\documentclass[a4paper,fleqn,usenatbib,useAMS]{mnras}

\usepackage{newtxtext,newtxmath}

\usepackage[T1]{fontenc}

\usepackage{graphicx}	
\usepackage{amsmath}	
\usepackage{amssymb}	

\newcommand{\snia}{SN~Ia}
\newcommand{\sneia}{SNe~Ia}

\newcommand{\opar}{\hbox{$^{\scriptsize \rm  o}$}}
\newcommand{\oparsub}[1]{\hbox{$^{\scriptsize \rm  o}_{#1}$}}

\let\ts=\thinspace
\newcommand{\one}{\ts {\sc i}}
\newcommand{\two}{\ts {\sc ii}}

\newcommand{\nifs}{\ensuremath{^{56}\rm{Ni}}}
\newcommand{\cofs}{\ensuremath{^{56}\rm{Co}}}
\newcommand{\fefs}{\ensuremath{^{56}\rm{Fe}}}
\newcommand{\msun}{\ensuremath{\rm{M}_{\odot}}}
\newcommand{\mtot}{\ensuremath{M_\mathrm{tot}}}
\newcommand{\mratio}{$M(\nifs)/$\mtot}

\newcommand{\kms}{\ensuremath{\rm{km\,s}^{-1}}}

\newcommand{\dmft}{\ensuremath{\Delta M_{15}(B)}}
\newcommand{\dmftbol}{\ensuremath{\Delta M_{15}(\rm{bol})}}
\newcommand{\mch}{\ensuremath{M_{\rm Ch}}}

\def\cmfgen{{\sc cmfgen}}

\title[Sub-\mch\ progenitors of low-luminosity \sneia]
{Evidence for sub-Chandrasekhar-mass progenitors of Type Ia supernovae
  at the faint end of the width-luminosity relation}
\author[St\'ephane Blondin et al.]
{
St\'ephane Blondin,$^{1}$\thanks{E-mail: stephane.blondin@lam.fr}
Luc Dessart,$^{2}$
D.~John Hillier,$^{3}$
and Alexei~M. Khokhlov$^{4}$\\
$^{1}$Aix Marseille Univ, CNRS, LAM, Laboratoire d'Astrophysique de
Marseille, Marseille, France\\ 
$^{2}$Unidad Mixta Internacional Franco-Chilena de Astronom\'ia (CNRS UMI 3386),
    Departamento de Astronom\'ia, Universidad de Chile,\\
    Camino El Observatorio 1515, Las Condes, Santiago, Chile\\
$^{3}$Department of Physics and Astronomy \& Pittsburgh Particle
Physics, Astrophysics, and Cosmology Center (PITT PACC), University of
Pittsburgh,\\ Pittsburgh, PA 15260, USA\\
$^{4}$Department of Astronomy \& Astrophysics, the Enrico Fermi
Institute, and the Computation Institute, The University of Chicago,
Chicago, IL 60637, USA
}

\date{Accepted 2016 September 28. Received 2016 September 02; in original form 2016 June 24}
\pubyear{2017}

\begin{document}
\label{firstpage}
\pagerange{\pageref{firstpage}--\pageref{lastpage}}
\maketitle


\begin{abstract}
The faster light-curve evolution of low-luminosity Type Ia supernovae
(\sneia) suggests that they could result from the explosion of white
dwarf (WD) progenitors below the Chandrasekhar mass (\mch). Here we
present 1D non-local thermodynamic equilibrium time-dependent
radiative transfer simulations of pure central detonations of
carbon-oxygen WDs with a mass ($\mtot$) between 0.88\,\msun\ and
1.15\,\msun, and a \nifs\ yield between 0.08\,\msun\ and 0.84\,\msun.
Their lower ejecta density compared to \mch\ models results in a more
rapid increase of the luminosity at early times and an enhanced
$\gamma$-ray escape fraction past maximum light. Consequently, their
bolometric light curves display shorter rise times and larger
post-maximum decline rates.  Moreover, the higher \mratio\ ratio at a
given \nifs\ mass enhances the temperature and ionization level in the
spectrum-formation region for the less luminous models, giving rise to
bluer colours at maximum light and a faster post-maximum evolution of
the $B-V$ colour.  For sub-\mch\ models fainter than $M_B \approx -18.5$\,mag
at peak, the greater bolometric decline and faster colour evolution
lead to a larger $B$-band post-maximum decline rate, \dmft.  In
particular, all of our previously-published \mch\ models (standard and
pulsational delayed detonations) are confined to $\dmft < 1.4$\,mag,
while the sub-\mch\ models with $\mtot \lesssim 1$\,\msun\ extend
beyond this limit to $\dmft \approx 1.65$\,mag for a peak $M_B \approx
-17$\,mag, in better agreement with the observed width-luminosity
relation (WLR).  Regardless of the precise ignition mechanism, these
simulations suggest that fast-declining \sneia\ at the faint end of the
WLR could result from the explosion of WDs whose mass is
significantly below the Chandrasekhar limit.
\end{abstract}



\section{Introduction}\label{sect:intro}

The finding that the peak magnitudes of Type Ia supernovae (\sneia)
are tightly correlated with their post-maximum decline rate in several
optical bands \citep{Pskovskii:1977,Phillips:1993} revolutionized the
use of these events as extra-galactic distance indicators and formed
the cornerstone for the discovery of dark energy \citep{R98,P99}.

Due to its determining role in observational cosmology, this so-called
width-luminosity relation (hereafter WLR when considering the
$B$-band) has received extensive theoretical attention.  The slower
post-maximum decline of \sneia\ with larger peak $M_B$ has been
thought to arise from an increase in opacity with \nifs\ mass, which
affects the timescale for the radiation to escape the ejecta
\citep[e.g.,][]{Hoeflich/etal:1996,Pinto/Eastman:2001,Mazzali/etal:2001}.
However, in contrast with observational studies that show a
significantly weaker relation between peak magnitude and post-maximum
decline rate for the integrated $(U)BVRI$ flux
\citep[e.g.,][]{Contardo/Leibundgut/Vacca:2000}, one would expect
this relation to also hold for the bolometric luminosity.

A later study by \cite{Kasen/Woosley:2007} showed that the decay
heating from \nifs\ affects the onset of the {\sc iii}$\rightarrow${\sc
  ii} recombination of iron-group elements (IGEs) in the ejecta, which
causes a faster colour evolution for the least luminous events (in
particular the $B-V$ colour), and hence a larger decline rate in $B$
relative to other bands. The WLR is then interpreted purely as a
colour effect related to the level of Fe\two/Co\two\ line blanketing,
independent of the evolution of the bolometric luminosity.

\begin{table*}
\footnotesize
\caption{Basic properties of the \snia\ models considered in this study, with
photometric quantities relevant to the width-luminosity relation.
}\label{tab:modelprop}
\begin{tabular}{l@{\hspace{1.4mm}}c@{\hspace{1.4mm}}c@{\hspace{1.4mm}}c@{\hspace{1.4mm}}c@{\hspace{1.4mm}}c@{\hspace{1.4mm}}c@{\hspace{1.4mm}}c@{\hspace{1.4mm}}c@{\hspace{1.4mm}}c@{\hspace{1.4mm}}c@{\hspace{1.4mm}}c@{\hspace{1.4mm}}c@{\hspace{1.4mm}}c}
\hline
\multicolumn{1}{l}{Model} & $M_{\mathrm{tot}}$ & $E_{\mathrm{kin}}$ & $M(\nifs)$ & $\varv_{99}(\nifs)$ & $\dot{e}_{\rm decay}$ & $t_{\mathrm{rise}}(\mathrm{bol})$ & $t_{\mathrm{rise}}(B)$ & $L_{\rm bol}$ & $M_B$ & $\Delta M_{15}(\mathrm{bol})$ & $\Delta M_{15}(B)$ & $(B-V)_{\mathrm{+0}}$ & $(B-V)_{\mathrm{+15}}$ \\
 & (M$_\odot$) & (erg) & (M$_\odot$) & (km s$^{-1}$) & (erg s$^{-1}$ g$^{-1}$) & (day) & (day) & (erg s$^{-1}$) & (mag) & (mag) & (mag) & (mag) & (mag) \\
\hline
\multicolumn{14}{c}{Standard Chandrasekhar-mass delayed-detonation models (DDC series)}\\
\hline
DDC0         & 1.41 & 1.56\,(51) & 0.86 & 1.29\,(4) & 6.32\,(9) & 16.7 & 17.7 & 1.85\,(43) & $-$19.65 & 0.74 & 0.74 & 0.02 & 0.20 \\
DDC6         & 1.41 & 1.50\,(51) & 0.72 & 1.20\,(4) & 5.41\,(9) & 16.8 & 18.0 & 1.57\,(43) & $-$19.48 & 0.72 & 0.83 & 0.04 & 0.36 \\
DDC10        & 1.41 & 1.48\,(51) & 0.62 & 1.15\,(4) & 4.63\,(9) & 17.1 & 17.7 & 1.38\,(43) & $-$19.33 & 0.74 & 0.92 & 0.04 & 0.50 \\
DDC15        & 1.41 & 1.47\,(51) & 0.51 & 1.12\,(4) & 3.72\,(9) & 17.6 & 17.1 & 1.14\,(43) & $-$19.15 & 0.72 & 1.03 & 0.04 & 0.65 \\
DDC17        & 1.41 & 1.41\,(51) & 0.41 & 1.08\,(4) & 2.85\,(9) & 18.6 & 17.2 & 9.10\,(42) & $-$18.93 & 0.67 & 1.27 & 0.04 & 0.90 \\
DDC20        & 1.41 & 1.38\,(51) & 0.30 & 1.03\,(4) & 2.12\,(9) & 18.7 & 17.2 & 6.65\,(42) & $-$18.44 & 0.58 & 1.38 & 0.28 & 1.14 \\
DDC22        & 1.41 & 1.30\,(51) & 0.21 & 9.80\,(3) & 1.40\,(9) & 19.6 & 17.5 & 4.47\,(42) & $-$17.62 & 0.54 & 1.22 & 0.73 & 1.34 \\
DDC25        & 1.41 & 1.18\,(51) & 0.12 & 8.56\,(3) & 7.37\,(8) & 21.0 & 19.8 & 2.62\,(42) & $-$16.44 & 0.62 & 1.01 & 1.31 & 1.50 \\
\hline
\multicolumn{14}{c}{Pulsating Chandrasekhar-mass delayed-detonation models (PDDEL series)}\\
\hline
PDDEL1       & 1.38 & 1.35\,(51) & 0.75 & 1.33\,(4) & 5.87\,(9) & 16.4 & 18.0 & 1.62\,(43) & $-$19.49 & 0.76 & 0.95 & 0.07 & 0.51 \\
PDDEL3       & 1.38 & 1.33\,(51) & 0.68 & 1.30\,(4) & 5.30\,(9) & 16.6 & 18.1 & 1.48\,(43) & $-$19.39 & 0.76 & 1.04 & 0.08 & 0.62 \\
PDDEL7       & 1.38 & 1.32\,(51) & 0.60 & 1.29\,(4) & 4.72\,(9) & 16.6 & 17.8 & 1.32\,(43) & $-$19.25 & 0.72 & 1.12 & 0.10 & 0.77 \\
PDDEL4       & 1.38 & 1.30\,(51) & 0.53 & 1.27\,(4) & 4.10\,(9) & 16.9 & 17.6 & 1.16\,(43) & $-$19.09 & 0.69 & 1.22 & 0.14 & 0.90 \\
PDDEL9       & 1.38 & 1.26\,(51) & 0.40 & 1.23\,(4) & 3.07\,(9) & 17.5 & 16.9 & 8.81\,(42) & $-$18.70 & 0.63 & 1.28 & 0.27 & 1.10 \\
PDDEL11      & 1.38 & 1.23\,(51) & 0.30 & 1.18\,(4) & 2.13\,(9) & 18.8 & 16.9 & 6.42\,(42) & $-$18.12 & 0.60 & 1.21 & 0.56 & 1.27 \\
PDDEL12      & 1.38 & 1.21\,(51) & 0.25 & 1.15\,(4) & 1.73\,(9) & 19.5 & 17.3 & 5.43\,(42) & $-$17.76 & 0.60 & 1.14 & 0.76 & 1.35 \\
\hline
\multicolumn{14}{c}{Sub-Chandrasekhar-mass models (SCH series)}\\
\hline
SCH7p0       & 1.15 & 1.40\,(51) & 0.84 & 1.51\,(4) & 7.91\,(9) & 15.6 & 15.8 & 1.85\,(43) & $-$19.64 & 0.85 & 0.79 & 0.06 & 0.12 \\
SCH6p5       & 1.13 & 1.35\,(51) & 0.77 & 1.44\,(4) & 7.43\,(9) & 15.8 & 16.2 & 1.71\,(43) & $-$19.58 & 0.86 & 0.81 & 0.05 & 0.14 \\
SCH6p0       & 1.10 & 1.30\,(51) & 0.70 & 1.39\,(4) & 6.91\,(9) & 16.0 & 16.7 & 1.57\,(43) & $-$19.50 & 0.86 & 0.83 & 0.05 & 0.17 \\
SCH5p5       & 1.08 & 1.23\,(51) & 0.63 & 1.32\,(4) & 6.31\,(9) & 16.3 & 17.3 & 1.42\,(43) & $-$19.39 & 0.87 & 0.90 & 0.06 & 0.26 \\
SCH5p0       & 1.05 & 1.16\,(51) & 0.55 & 1.25\,(4) & 5.67\,(9) & 16.4 & 17.3 & 1.25\,(43) & $-$19.25 & 0.87 & 0.97 & 0.06 & 0.39 \\
SCH4p5       & 1.03 & 1.09\,(51) & 0.46 & 1.20\,(4) & 4.95\,(9) & 16.6 & 17.5 & 1.08\,(43) & $-$19.09 & 0.86 & 1.10 & 0.07 & 0.57 \\
SCH4p0       & 1.00 & 1.03\,(51) & 0.38 & 1.16\,(4) & 4.23\,(9) & 16.5 & 17.3 & 9.01\,(42) & $-$18.88 & 0.84 & 1.24 & 0.11 & 0.74 \\
SCH3p5       & 0.98 & 9.71\,(50) & 0.30 & 1.14\,(4) & 3.50\,(9) & 16.4 & 16.5 & 7.34\,(42) & $-$18.62 & 0.81 & 1.38 & 0.16 & 0.91 \\
SCH3p0       & 0.95 & 9.19\,(50) & 0.23 & 1.11\,(4) & 2.73\,(9) & 16.5 & 15.2 & 5.76\,(42) & $-$18.30 & 0.80 & 1.47 & 0.22 & 1.07 \\
SCH2p5       & 0.93 & 8.67\,(50) & 0.17 & 1.08\,(4) & 2.08\,(9) & 16.4 & 14.9 & 4.36\,(42) & $-$17.86 & 0.80 & 1.60 & 0.39 & 1.22 \\
SCH2p0       & 0.90 & 8.14\,(50) & 0.12 & 1.05\,(4) & 1.52\,(9) & 15.9 & 14.6 & 3.17\,(42) & $-$17.27 & 0.85 & 1.64 & 0.67 & 1.31 \\
SCH1p5       & 0.88 & 7.59\,(50) & 0.08 & 1.02\,(4) & 1.09\,(9) & 15.4 & 13.4 & 2.26\,(42) & $-$16.64 & 0.98 & 1.52 & 0.89 & 1.34 \\
\hline
\end{tabular}
\flushleft
{\bf Notes:}
Numbers in parenthesis correspond to powers of ten.
The \nifs\ mass corresponds to $t\approx 0$, and
$\varv_{99}(\nifs)$ is the velocity of the ejecta shell that bounds 99\% of the total \nifs\ mass.
$\dot{e}_{\rm decay}$ is the specific heating rate at bolometric maximum, corresponding to 
the instantaneous rate of decay energy actually deposited in the ejecta at this time ($L_{\rm decay}$) divided by the total mass (\mtot).
The last two columns give the $B-V$ colour at $B$-band maximum and 15 days later.

\end{table*}

The mass of the exploding WD (hereafter $M_{\rm tot}$) is known to
affect the bolometric evolution around maximum light
\citep{Pinto/Eastman:2000a}. However, most theoretical studies of the
WLR have either focused exclusively on Chandrasekhar-mass (\mch)
progenitors
\citep[e.g.,][]{Pinto/Eastman:2001,Mazzali/etal:2001,Hoeflich/etal:2002,Kasen/Woosley:2007},
or excluded sub-\mch\ double-detonation models based on their apparent
colour mismatch \citep[see, e.g.,][]{Hoeflich/etal:1996}.  A generic
problem of these sub-\mch\ models resides in the composition of the
detonated He shell, which is needed to trigger a secondary detonation
in the carbon-oxygen (C-O) core. The synthesis of \nifs\ and other
IGEs in this external shell causes discrepant colours and spectra at early
times \citep{Nugent/etal:1997,Kromer/etal:2010}.  Nonetheless,
\cite{Sim/etal:2010} have shown that pure central detonations in
single sub-\mch\ WDs without an outer He shell are able to reproduce
the overall trend of the observed WLR, although it is still unclear
whether sub-\mch\ models fare better than the standard \mch\ scenario,
in particular for low-luminosity \sneia. However, observational
studies have shown that the narrow light curves of fast-declining
\sneia\ likely result from sub-\mch\ ejecta, and that the WLR
is naturally explained as a relation between ejected mass and
\nifs\ mass \citep{Scalzo/etal:2014a,Scalzo/etal:2014b}.

While the WLR imposes very strict constraints on \snia\ models, its
reproduction by no means guarantees a high fidelity to observed events
\citep[see, e.g.,][]{Blondin/etal:2011b}.  Moreover, the emphasis on
the importance of the ionization balance in \snia\ ejecta to explain
the WLR highlights the need for an accurate determination of the
atomic level populations via a solution to the statistical equilibrium
equations, i.e., a fully non-local-thermodynamic equilibrium (non-LTE)
solution to the radiative transfer problem, often approximated in
previous studies.  Here we present 1D non-LTE time-dependent radiative
transfer simulations of sub-\mch\ models for
\sneia\ (Section~\ref{sect:models}), performed using the
\cmfgen\ radiative-transfer code \citep{Hillier/Dessart:2012}.

To better illustrate the impact of a low ejecta mass on the
  radiative display, we confront these sub-\mch\ models to our
  previously-published \mch\ models,  which include ``standard''
\mch\ delayed-detonations (DDC series; \citealt{Blondin/etal:2013})
and pulsational \mch\ delayed-detonations (PDDEL series;
\citealt{D14_PDD}). Despite the one-dimensional treatment of the
explosion, these \mch\ models were found to capture the most salient
features of more elaborate multi-dimensional simulations, in particular the
asymptotic kinetic energy and chemical stratification \citep[see,
  e.g.,][]{Seitenzahl/etal:2013}. More 
importantly, they provided an excellent match to the observed
properties of both normal and low-luminosity \sneia\ at maximum light
\citep{Blondin/etal:2013}, and to the spectral evolution of normal
events out to $\lesssim 100$\,d past explosion (SN~2002bo,
\citealt{Blondin/etal:2015}; SN~2005cf, \citealt{D14_Tech}; SN~2011fe,
\citealt{D14_PDD}).

We start by analyzing the bolometric evolution of the
sub-\mch\ models, highlighting the differences in rise time and
post-maximum decline compared to \mch\ models at a given peak
luminosity, i.e.,  similar \nifs\ mass but different
WD mass, \mtot\ (Section~\ref{sect:bol}).
We then discuss the impact of a sub-\mch\ ejecta on the
maximum-light $B-V$ colour (Section~\ref{sect:colmax}), and on its
post-maximum evolution for low-luminosity
\sneia\ (Section~\ref{sect:colevol}).
Last, we assess the merits
of sub-\mch\ progenitors in reproducing the observed WLR all the way
to the faint end (Section~\ref{sect:wlr}).  We discuss possible
progenitor scenarios leading to the detonation of a sub-\mch\ WD in
Section~\ref{sect:ccl}, followed by our conclusions.


\section{Sub-Chandrasekhar-mass models}\label{sect:models}

The sub-\mch\ models studied here correspond to pure central
detonations of WDs with a mass ranging between 0.88\,\msun\ and
1.15\,\msun\ (Table~\ref{tab:modelprop}).
As such they are similar to the sub-\mch\ models of
  \cite{Sim/etal:2010}, albeit with a different numerical treatment
  and radiative-transfer post-processing (see below).
The initial WD is assumed to
be in hydrostatic equilibrium, and is composed of equal amounts of
$^{12}$C and $^{16}$O by mass, with traces of $^{22}$Ne and solar
composition for all other isotopes. Importantly, we do not
consider the presence of an external He shell, required in the
double-detonation scenario to trigger a detonation in the C-O
core. Despite the artificial nature of this setup, the models are
nonetheless intrinsically coherent and enable us to qualitatively
assess the impact of a lower ejecta mass on the radiative display.
The 1D hydrodynamical treatment of the explosion phase is analogous to
that used in our previous \snia\ studies based on
\mch\ delayed-detonation models \citep[see][their
    section~2]{Blondin/etal:2013}. The calculation is carried out until the
ejecta mass shells reach a ballistic regime (homologous expansion),
less than $\sim 20$\,s past explosion. We smooth sharp variations in
the composition and density profiles of the hydrodynamical input
through convolution with a Gaussian kernel of width $\sigma =
800$\,\kms\ \citep[see also][]{Blondin/etal:2015}.

One important difference is the absence of an initial deflagration
phase, required to pre-expand the WD in \mch\ models and hence
synthesize intermediate-mass elements during the secondary detonation
phase. In sub-\mch\ models the burning can proceed directly in the
form of a detonation wave, since the lower WD density ensures the
burning will be incomplete in the outer ejecta layers with no prior
pre-expansion. As a result, the initial WD structure sets the density
at which the C-O fuel is burned, which is lower for lower WD
masses. In particular, the amount of \nifs\ synthesized in the
explosion is directly proportional to the WD mass, as is the
asymptotic kinetic energy (see Table~\ref{tab:modelprop}). In our
sub-\mch\ model series (SCH), the initial range of WD masses
(0.88--1.15\,\msun) translates into a \nifs\ yield between
0.08\,\msun\ and 0.84\,\msun. 

The lower WD density also limits the production of
  neutron-rich stable isotopes of IGEs (e.g., $^{54}$Fe, $^{58}$Ni) in
  the inner ejecta, while 
  these are more abundantly produced during the deflagration phase of
  \mch\ models. In the outer ejecta layers,
  however, the burning proceeds at similar densities in
  both the sub-\mch\ and \mch\ models considered here, resulting in
  similar abundance profiles. At a given \nifs\ yield, the
  sub-\mch\ models can thus be considered as low-mass analogs of the
  \mch\ models when discussing their radiative properties at early
  times, including the maximum-light phase discussed in the present
  paper.

The long-term evolution is computed with the 1D, time-dependent,
non-LTE radiative-transfer code
\cmfgen\ \citep{Hillier/Dessart:2012,D14_Tech}, which includes the
treatment of non-local energy deposition and non-thermal processes
\citep[e.g.,][]{D14_CoIII}.
Apart from the two-step \nifs$\rightarrow$\cofs$\rightarrow$\fefs\ 
decay chain, we treat eight additional two-step decay chains associated with
$^{37}$K, $^{44}$Ti, $^{48}$Cr, $^{49}$Cr, $^{51}$Mn, $^{52}$Fe,
$^{55}$Co, $^{57}$Ni, and a further six one-step decay chains associated
with $^{41}$Ar, $^{42}$K, $^{43}$K, $^{43}$Sc, $^{47}$Sc, $^{61}$Co
\citep[see][]{D14_Tech}.  
All atom/ion level populations are
determined explicitly through a solution of the time-dependent
statistical equilibrium equations, which are coupled to the energy and
radiative-transfer equations. We consider the following ions:
C\one--{\sc iii}, O\one--{\sc iii}, Ne\one--{\sc iii}, Na\one,
Mg\two--{\sc iii}, Al\two--{\sc iii}, Si\two--{\sc iv}, S\two--{\sc
  iv}, Ar\one--{\sc iii}, Ca\two--{\sc iv}, Sc\two--{\sc iii},
Ti\two--{\sc iii}, Cr\two--{\sc iv}, Mn\two--{\sc iii}, Fe\one--{\sc
  vii}, Co\two--{\sc vii}, and Ni\two--{\sc vii}. The
number of levels considered for each ion is given in
Appendix~\ref{sect:atom}, Table~\ref{tab_atom_big}.
We assume a constant effective Doppler width (including
  both thermal and turbulent velocities) of 50\,\kms\ for our
  line-absorption profiles. While this value largely overestimates the
  true effective Doppler width (likely a few \kms\ at most) and hence
  the ejecta opacity, we have run extensive tests imposing an
  effective Doppler width of 10\,\kms\ and found a negligible impact
  (at the few per cent level) on the predicted rise times and
  post-maximum decline rates.  The output of such calculations are
light curves and spectra that can be directly confronted to
\snia\ observations. Our assumption of spherical symmetry allows for
the most elaborate radiative-transfer treatment to date of such
sub-\mch\ models.

In what follows we focus on the properties most relevant to the WLR,
while the full SCH model series will be presented in more detail in a
forthcoming paper.


\section{Bolometric rise and post-maximum decline}\label{sect:bol}

\begin{figure}
\centering
\includegraphics{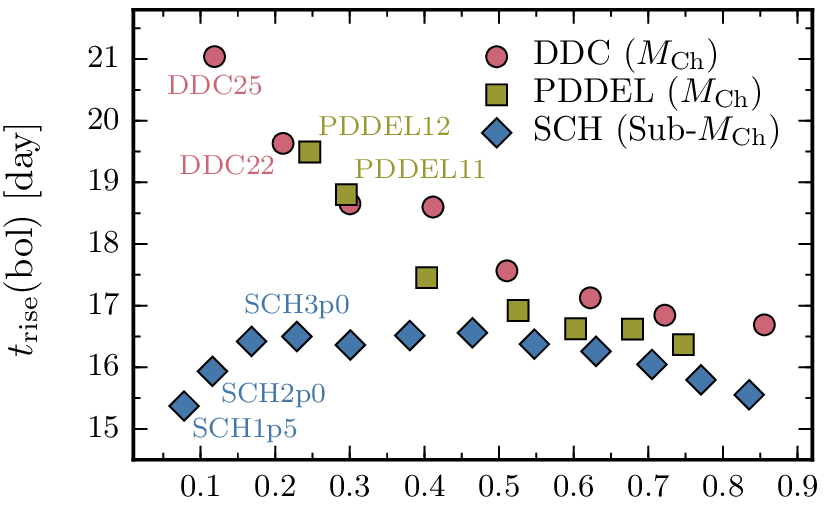}\vspace{-.3cm} 
\includegraphics{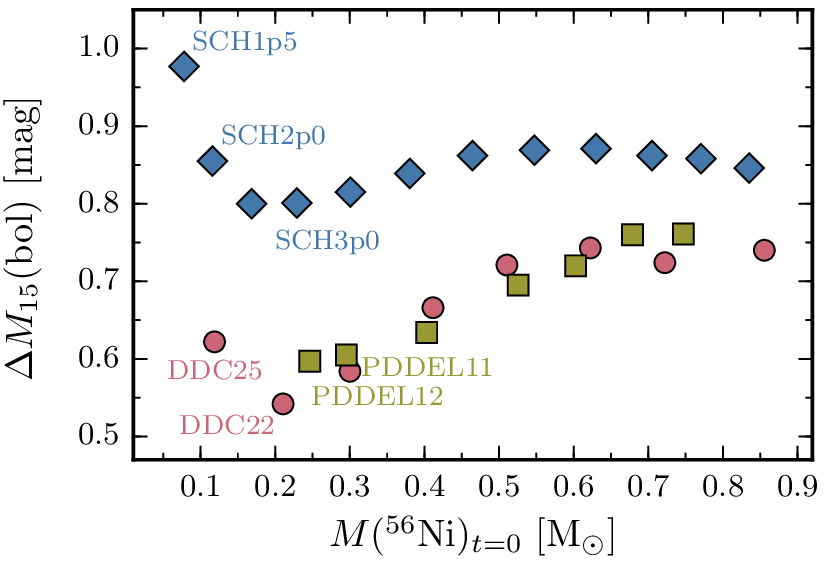}
\caption{\label{fig:trisedm15bol} Bolometric rise time (top) and
  decline-rate \dmftbol\ (bottom) versus \nifs\ mass for the
  \mch\ (DDC series, circles; PDDEL series, squares) and
  sub-\mch\ models (diamonds).  }
\end{figure}

\cite{Pinto/Eastman:2000a} have emphasized the impact of the total
mass\footnote{In this paper we only consider explosions that
  completely unbind the WD, hence the ejecta mass is equal to the
  WD mass.} on the bolometric light curves of \sneia, through its
effect on the ejecta density and $\gamma$-ray escape. A decrease in
\mtot\ results in a lower column depth which favours a more rapid
release of radiation at early times and an enhanced $\gamma$-ray
escape fraction past maximum light, leading to a shorter rise time and
a faster post-maximum decline.  This is indeed the case for the
sub-\mch\ models presented here, which display systematically shorter
bolometric rise times than the \mch\ models for a given \nifs\ mass
(Fig.~\ref{fig:trisedm15bol}, top panel).

Due to the shorter rise time, the instantaneous rate of energy deposition by
radioactive decays ($L_{\rm decay}$) at maximum light is larger 
(as is the peak luminosity, $L_{\rm bol}$), 
resulting in a more efficient heating of the ejecta compared to a
\mch\ model with the same \nifs\ mass. The lower \mtot\ exacerbates
this effect, as the {\it specific} heating rate at maximum light
($\equiv L_{\rm decay}/$\mtot, noted $\dot{e}_{\rm decay}$ in
Table~\ref{tab:modelprop}) is even larger. This latter parameter is
key in explaining the bluer maximum-light colours of the
sub-\mch\ models (see Section~\ref{sect:colmax}).

Moreover, all of the sub-\mch\ models have a bolometric decline rate
$\dmftbol \ge 0.80$\,mag (and up to 0.98\,mag), where the \mch\ models
are confined to smaller values (0.54--0.76\,mag; see
Fig.~\ref{fig:trisedm15bol}, bottom panel). This larger bolometric
decline for the sub-\mch\ models will naturally affect all photometric
bands, including the $B$-band on which the WLR is based
(Section~\ref{sect:wlr}).

That said, both the \mch\ and sub-\mch\ models share a weak dependence
of their post-maximum bolometric decline rate on the \nifs\ mass: the
variation in \dmftbol\ is less than $\sim0.2$\,mag across a factor
$\sim$7--10 variation in \nifs\ mass. When excluding the faintest
sub-\mch\ model (SCH1p5), the difference in \dmftbol\ drops to a mere
0.07\,mag for the SCH series. This contrasts with the far stronger
dependence of the post-maximum $B$-band decline rate \dmft\ on
\nifs\ mass (and hence peak luminosity; see
Table~\ref{tab:modelprop}), and lends support to the interpretation of
the WLR as a colour effect as opposed to an increase of the diffusion
time with \nifs\ mass \citep[see][]{Kasen/Woosley:2007}.

In fact, the bolometric evolution more naturally follows an opposite
trend to the ($B$-band) WLR, the more luminous models declining more
rapidly (this is especially true in the \nifs\ mass range
0.2--0.6\,\msun; see Fig.~\ref{fig:trisedm15bol}). The bolometric
decline rate is related to the magnitude and rate of change of the
$\gamma$-ray escape fraction past maximum light, which is larger for
higher \nifs\ mass \citep[see][]{Stritzinger/etal:2006a}. This is in
part due to the larger outward extent of the \nifs\ distribution (see
the $\varv_{99}(\nifs)$ column in Table~\ref{tab:modelprop}) which
favours the earlier and more rapid escape of $\gamma$-rays, thereby
limiting the amount of decay energy actually deposited in the ejecta
\citep[see also][]{Pinto/Eastman:2001}. This effect is modulated by
the release of stored radiation past maximum light, which leads to a
secondary bolometric maximum whose timing can affect the value of
\dmftbol.


\section{Maximum-light colours}\label{sect:colmax}

\begin{figure}
\centering \includegraphics{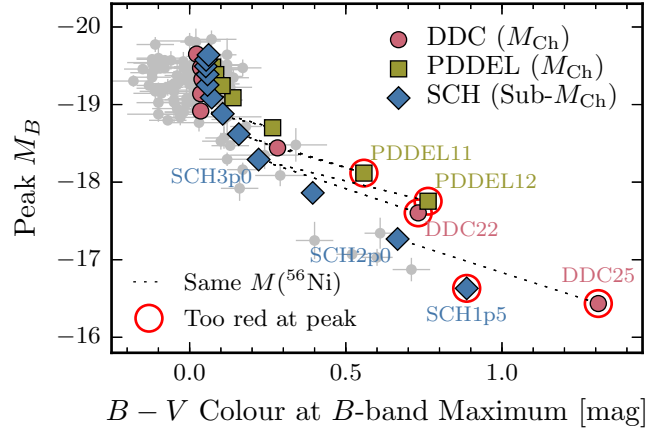}
\caption{\label{fig:bmvp0} Peak $B$-band magnitude versus $B-V$ colour
  at $B$-band maximum for the \mch\ (DDC series, circles; PDDEL
  series, squares) and sub-$M_{\rm Ch}$ models (diamonds). Models with
  a similar \nifs\ mass (within $\pm0.02$\,\msun) are connected with a
  dotted line.  Also shown are measurements taken from
  {\protect\cite{Hicken/etal:2009a}} [grey points]. Models that have a
  too red $B-V$ colour at $B$-band maximum are highlighted with a red
  circle.  }
\end{figure}

Type Ia supernovae obey a strong relation between their peak $M_B$
and their $B-V$ colour at maximum light, brighter events displaying
bluer colours \citep[e.g.,][]{Tripp:1998}. This relation simply
reflects the larger magnitude of decay heating from \nifs\ for more
luminous events, and holds for both \mch\ and sub-\mch\ models
(Fig.~\ref{fig:bmvp0}).

Nonetheless, one would expect the greater {\it specific} heating rate
($\dot{e}_{\rm decay}$; see previous section) for the sub-\mch\ models
at a given \nifs\ mass to result in even bluer colours.  As seen in
Fig.~\ref{fig:bmvp0}, this effect is present and more pronounced at
the faint end (up to $\sim0.6$\,mag bluer $B-V$ colour at a given
\nifs\ mass), where the
relative difference in $\dot{e}_{\rm decay}$ is largest (see
Table~\ref{tab:modelprop}). In contrast, all models brighter than
$M_B=-19$\,mag at peak have very similar $B-V$ colours at maximum
($\lesssim 0.03$\,mag standard deviation).

The sub-\mch\ models follow the observed relation all the way to the
faint end (peak $M_B \gtrsim -17$\,mag), while the \mch\ models are
systematically too red at a given peak $M_B$. This is also the case
for the least luminous sub-\mch\ model (SCH1p5), in which the \nifs\ mass is
too low ($<0.08$\,\msun, where all the other models have $\gtrsim
0.12$\,\msun\ of \nifs; see Table~\ref{tab:modelprop}) to efficiently
heat the spectrum-formation layers at maximum light. In fact, the only
model to have an even redder $B-V$ colour at maximum is the
\mch\ model DDC25, which has the lowest specific heating rate
of all ($\dot{e}_{\rm decay} < 10^9$\,erg\,s$^{-1}$\,g$^{-1}$).


\section{The post-maximum colour evolution of low-luminosity
  \sneia}\label{sect:colevol} 

\begin{figure}
\centering
\includegraphics{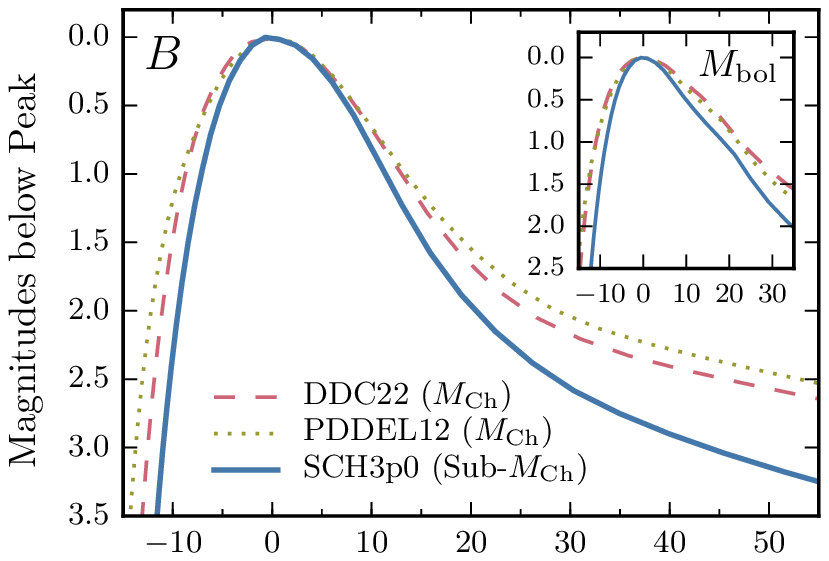}\vspace{.2cm}
\includegraphics{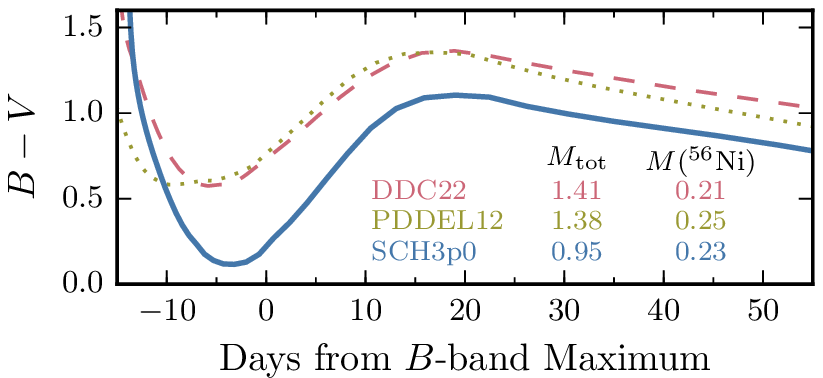}
\caption{\label{fig:mbnormbmv} Normalized $B$-band light curves (top)
  and $B-V$ colour evolution (bottom) for the standard
  \mch\ delayed-detonation model DDC22 (dashed line), the pulsational
  \mch\ delayed-detonation model PDDEL12 (dotted line), and the
  sub-\mch\ model SCH3p0 ($M_{\rm tot}=0.95\,\msun$; solid line), all
  of which have a similar \nifs\ mass of $0.23\pm0.02$\,\msun. The
  inset shows the normalized bolometric light curves, where the time
  axis now corresponds to days from bolometric maximum.  }
\end{figure}

We focus on three models with similar \nifs\ mass (within $\pm
0.02$\,\msun\ of 0.23\,\msun) to better illustrate the impact of
\mtot\ on the post-maximum colour evolution of low-luminosity \sneia:
the standard \mch\ delayed-detonation model DDC22, the pulsational
\mch\ delayed-detonation model PDDEL12, and the sub-\mch\ model
SCH3p0, corresponding to the pure central detonation of a 
0.95\,\msun\ C-O WD. Their $B$-band light curves (normalized to the same
peak magnitude) and $B-V$ colour evolution is shown in
Fig.~\ref{fig:mbnormbmv}. One clearly notes the faster evolution of
the $B$-band light curve for the sub-\mch\ model (in part due to the
faster bolometric evolution; see Fig.~\ref{fig:mbnormbmv} inset), with
a $\sim 2$\,d shorter rise time and a $> 0.3$\,mag larger \dmft\ [see
  Table~\ref{tab:modelprop}], accompanied by a faster evolution of the
$B-V$ colour around maximum light. This results in a gradual decrease
in the difference in $B-V$ colour compared to the \mch\ models, from
$>0.5$\,mag at $B$-band maximum to $<0.3$\,mag 15 days later.

The greater specific heating rate of the sub-\mch\ model
($\dot{e}_{\rm decay}$; see Section~\ref{sect:bol}) results in a
relatively modest increase of the temperature in the spectrum-formation
region at maximum light compared to the \mch\ models (on the order of
10--15\%). Combined with the lower ejecta density, this slightly
larger temperature is nonetheless sufficient to induce a change in the
mean ionization state of the gas, enhancing the {\sc iii}/{\sc ii}
ionization ratio of IGEs (Sc, Ti, Cr, Fe, and Co) that would otherwise
effectively block the flux in the $B$-band and emit in redder bands in
their once-ionized state (especially Fe\two; see
Appendix~\ref{sect:ladder}).

\begin{figure}
\centering
\includegraphics{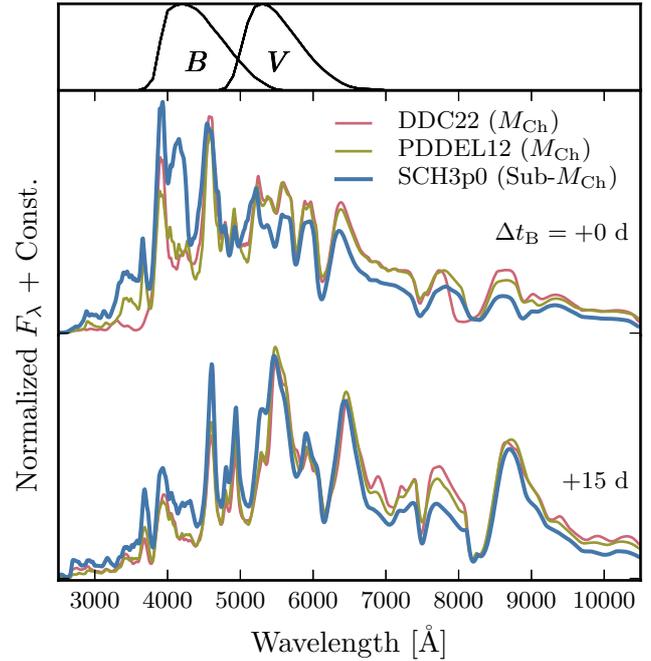}
\caption{\label{fig:specp0p15} Optical spectra at $B$-band maximum and
  15 days later, for the standard \mch\ delayed-detonation model DDC22
  (red line), the pulsational \mch\ delayed-detonation model PDDEL12
  (yellow line), and the sub-\mch\ model SCH3p0 ($M_{\rm
    tot}=0.95\,\msun$; blue line), all of which have a similar
  \nifs\ mass of $0.23\pm0.02$\,\msun. The spectra have been
  normalized to their mean flux in the range 3000--10000\,\AA.  The
  upper panel shows the normalized transmission curves for the $B$ and
  $V$ filters.}
\end{figure}

The bluer $B-V$ colour of the sub-\mch\ model thus results both from
an overall shift of the SED to the blue (akin to a purely thermal
effect) and a modulation of line-blanketing from changes in
ionization. Both effects are clearly visible in the maximum-light
spectra shown in Fig.~\ref{fig:specp0p15}. By 15 days past $B$-band
maximum, the temperature in the spectrum-formation region of the
sub-\mch\ model has decreased sufficiently to enhance the absorption
by singly-ionized IGEs to a level comparable to that in the
\mch\ models.  The differences in $B$-band flux as well as in the
overall SED shape are greatly reduced at this phase, although the
sub-\mch\ model remains slightly bluer, thereby limiting the $B$-band
post-maximum decline rate.


\section{The width-luminosity relation}\label{sect:wlr}

\begin{figure*}
\centering \includegraphics{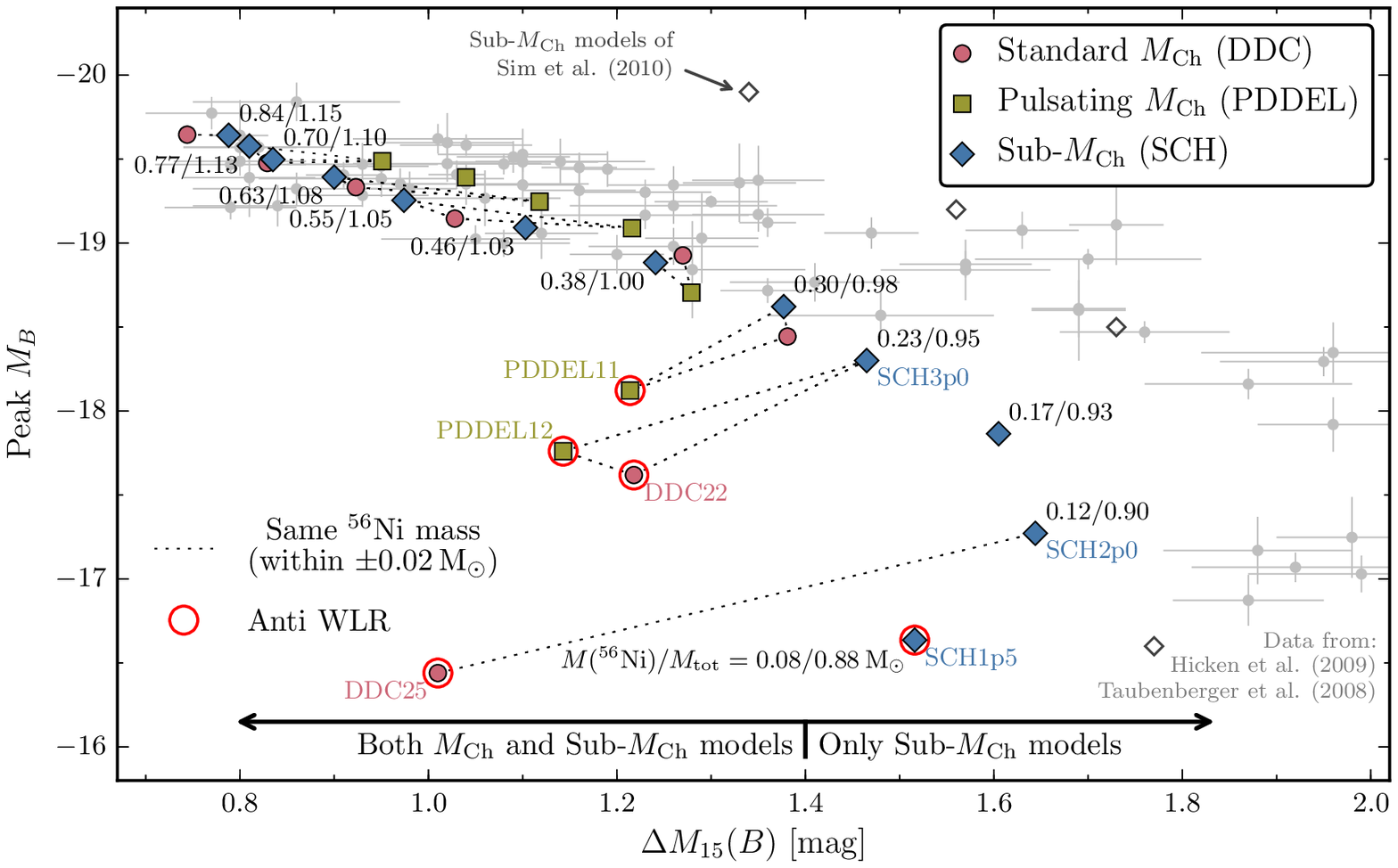}
\caption{\label{fig:wlrb} Width-luminosity relation for the \mch\ (DDC
  series, circles; PDDEL series, squares) and sub-$M_{\rm Ch}$ models
  (filled diamonds; the numbers associated with each symbol corresponds to
  the \nifs\ and progenitor WD mass, respectively). Models with a
  similar \nifs\ mass (within $\pm0.02$\,\msun) are connected with a
  dotted line.  Also shown are the pure C-O sub-\mch\ models of
    {\protect\cite{Sim/etal:2010}} [open diamonds], as well as
  measurements taken from 
  {\protect\cite{Hicken/etal:2009a}} and
  {\protect\cite{Taubenberger/etal:2008}} [grey points]. Models that
  have transitioned to an {\it anti} WLR are highlighted with a red
  circle.  These correspond to the same models that have a too red
  $B-V$ colour at maximum light in Fig.~\ref{fig:bmvp0}.  Only the
  sub-\mch\ models yield $B$-band light curves with a decline-rate
  $\dmft>1.4$\,mag.  }
\end{figure*}

The WLR for the \mch\ and sub-\mch\ models is shown in
Fig.~\ref{fig:wlrb}.  At the bright end (peak $M_B \lesssim
-18.5$\,mag), both \mch\ and sub-\mch\ models follow the observed WLR
(grey data points in Fig.~\ref{fig:wlrb}), albeit with slightly
steeper slopes (similar for the DDC and SCH series, and somewhat
shallower for the PDD series).  At the faint end (peak $M_B \gtrsim
-18.5$\,mag), the \mch\ models display a turnover to an {\it anti}
WLR, reaching a maximum \dmft\ value of $\sim 1.4$\,mag (DDC series)
and $\sim 1.3$\,mag (PDDEL series). These models correspond to the
same ones that have a too red $B-V$ colour at maximum light in
Fig.~\ref{fig:bmvp0} (highlighted with red circles in both figures),
due to the less efficient heating of the spectrum-formation region
from the lower $\dot{e}_{\rm decay}$ (Section~\ref{sect:colmax}).
Combined with a modest post-maximum evolution of their $B-V$ colour,
the resulting $B$-band decline rate is then too low for the
\mch\ models to reproduce the faint end of the WLR despite their low
luminosity.

In the sub-\mch\ models, the more rapid colour evolution around
maximum light combined with a larger bolometric decline rate result in
a higher \dmft\ for the same \nifs\ mass.  Consequently, only the
sub-\mch\ models extend beyond $\dmft = 1.4$\,mag. This confirms the
WD mass as an essential parameter in producing rapidly-declining
\sneia\ at the faint end of the WLR.  In particular, the
sub-\mch\ model SCH2p0 ($M_{\rm tot}=0.90$\,\msun) has a
$\sim 0.6$\,mag larger \dmft\ value than the \mch\ model with the same
\nifs\ mass, DDC25 ($M(\nifs)=0.12$\,\msun\ for both).

We note, however, that the sub-\mch\ models also transition to an
anti-WLR, albeit at much fainter magnitudes (peak $M_B \gtrsim
-17$\,mag), and that they fail to reach the highest observed
\dmft\ values.  We speculate that small variations in the
\mratio\ ratio or in the \nifs\ distribution within the ejecta
(due to explosion asymmetries or large-scale mixing) could 
enhance the rate of $\gamma$-ray escape and lead to a better agreement
with the data in this respect. We will discuss this in more detail in
a companion paper in which we compare the SCH2p0 model to the
low-luminosity SN~1999by.

The sub-\mch\ models of \cite{Sim/etal:2010} yield systematically
larger \dmft\ values compared to our SCH models at a given peak $M_B$,
by up to $\sim0.6$\,mag for their two most luminous models and
$\sim0.3$\,mag for their two least luminous models
(Fig.~\ref{fig:wlrb}). Their most luminous model ($M(\nifs)/M_{\rm
  tot} = 0.81/1.15$\,\msun, similar to our SCH7p0 model) declines too
slowly for its luminosity, with $\dmft\approx 1.3$\,mag for a peak
$M_B\approx-20$, where \sneia\ of comparable luminosity (and our
SCH7p0 model) have $\dmft\approx0.8$\,mag. Their least luminous model,
however, is in better agreement with the observed WLR, with
$\dmft\approx1.8$\,mag where our SCH1p5 model only has
$\dmft\approx1.5$\,mag for the same peak
$M_B\approx-16.6$. Differences in the nucleosynthetic post-processing
of the explosion models and in the radiative-transfer treatment are
the likely causes of these offsets (Sim, priv. comm.).

As was the case for the bolometric light curves
(Section~\ref{sect:bol}), the faster $B$-band post-maximum decline of
the sub-\mch\ models is associated with a shorter $B$-band rise time,
as observed for low-luminosity
\sneia\ \citep{Ganeshalingam/etal:2011}. To match the current practice
of observers, the time reference is set to maximum light. Studies of
the WLR thus ignore the scatter in rise times for \sneia\ with
different peak $M_B$. While determining the precise time of explosion
is observationally complicated, the recent examples of SN~2011fe
\citep{Nugent/etal:2011}, SN~2013dy \citep{Zheng/etal:2013}, and three
\snia\ candidates discovered during the Kepler mission
\citep{Olling/etal:2015} illustrate the potential of high-cadence
searches to constrain the rise time. Future studies of the WLR will
then need to consider the pre-maximum rise as well as the post-maximum
decline when distinguishing between fast and slow light-curve
evolution.


\section{Discussion and Conclusions}\label{sect:ccl}

The predicted observational signatures of pure central detonations in
sub-\mch\ WDs corroborate a number of distinctive properties of
low-luminosity \sneia.  At a given \nifs\ mass, the lower ejecta mass
compared to \mch\ models results in a faster bolometric
evolution around maximum light, which affects all photometric
bands. Furthermore, the larger \mratio\ ratio leads to a bluer $B-V$
colour at maximum light and a higher post-maximum $B$-band decline
rate, \dmft, in better agreement with the observed WLR. Conversely,
the lower {\it specific} heating rate in the \mch\ models
considered here results in a $B-V$ colour at maximum that
is too red compared to observations (through excess absorption by
singly-ionized IGEs and weaker emission in the blue), and hence a more
modest post-maximum colour evolution that imposes an upper limit on
the \dmft\ value.

Even then, the sub-\mch\ models fail to match the fastest decline
rates observed for low-luminosity, 91bg-like \sneia. Variations in the
\mratio\ ratio or in the \nifs\ distribution within the ejecta could
enhance the rate of $\gamma$-ray escape and result in a more efficient
cooling of the spectrum-formation region at these
phases. \cite{Sim/etal:2010} also find that a modest change in the
initial WD composition (replacing 7.5\% by mass of $^{12}$C with
$^{22}$Ne) has a non-negligible impact on the resulting \dmft\ value.
We note that a similar \mratio\ ratio as our low-luminosity
sub-\mch\ models can in principle be achieved in the context of
\mch\ models if the explosion is too weak to completely unbind the WD,
leaving behind a bound remnant \citep[see][]{Fink/etal:2014}.

The sub-\mch\ (and \mch) models studied here remain somewhat
artificial for genuine quantitative predictions to be made, yet the
level of sophistication enabled by the 1D treatment of the radiative
transfer enables us to qualitatively assess the impact of a lower
ejecta mass on the ionization balance for a given \nifs\ mass, which
is crucial in explaining the observed WLR. Ultimately, our results
will need to be confirmed by more realistic, multi-dimensional models,
although radiative-transfer simulations with the same level of
sophistication as ours are currently not feasible in 3D \citep[see,
  e.g.,][]{Sim/etal:2013}.

But do such low-mass explosions occur in Nature? Detonations of
sub-\mch\ WDs require an external trigger, usually via thermal
instabilities in a layer of accreted He (double detonation mechanism;
see, e.g., \citealt{Woosley/Weaver:1994}). This layer can detonate at
a lower mass than previously thought \citep{Bildsten/etal:2007}, but
the resulting IGEs synthesized at high velocities cause discrepant
colours and spectra at early times \citep{Kromer/etal:2010}.  More
recent calculations by \cite{Shen/Moore:2014}, however, show that a
detonation in a 0.005\,\msun\ He shell on a 1.0\,\msun\ WD only
produces $^{28}$Si and $^{4}$He, and that significant
$^{44}$Ti/$^{48}$Cr production only occurs for shells more massive
than $\sim 0.01$\,\msun. Double detonations thus remain an attractive
scenario for fast-declining \sneia\ for which empirical determinations
of the ejecta mass indicate sub-\mch\ progenitors
\citep{Scalzo/etal:2014a,Scalzo/etal:2014b}.

Likewise, He-ignited violent mergers
\citep{Pakmor/etal:2013}, where the dynamical detonation of
0.01\msun\ of He is sufficient to trigger a detonation in the C-O
core, also synthesize very little IGEs in the outer layers
($\sim10^{-8}$\,\msun).
In particular, \cite{Pakmor/etal:2013} argue
that low-luminosity \sneia\ result from the violent merger of a C-O+He
WD system, which is less massive (and hence results in a more rapid
light-curve evolution) than double C-O WD mergers. The predicted rate
of such systems can in principle account for the observed \snia\ rate
\citep{Ruiter/etal:2011}, and their long delay times would corroborate
the association of low-luminosity \sneia\ with older stellar populations
\citep[e.g.,][]{Howell:2001}.

Another possibility, suggested by \cite{vanKerkwijk/etal:2010},
involves the merger of two low-mass WDs whose combined mass is less
than the Chandrasekhar limit (a $0.6+0.6$\,\msun\ merger in their
paper).  The explosion does not occur during the merger event, but at
a later time via compressional heating through accretion of the thick
disk from the merger remnant
\citep[see][]{Loren-Aguilar/etal:2009}. However, this scenario is
unlikely to hold for a combined merger mass $\lesssim 1.2$\,\msun,
given the high temperature and density conditions required to ignite
carbon (van Kerkwijk, priv. comm.), and hence would fail to reproduce
the fast-declining \sneia\ at the faint end of the WLR, which we argue
result from ejecta masses $\lesssim 1$\,\msun.
In this respect, head-on collisions of two 0.5\,\msun\ WDs
  represent an interesting alternative: \cite{Kushnir/etal:2013} find that
such a model yields 0.11\,\msun\ of \nifs, comparable to our SCH2p0
model. However, most studies of this progenitor channel conclude that
WD collisions can account for at most a few per cent of the observed
\snia\ rate \citep[see, e.g.,][]{Papish/Perets:2016}.

Regardless of the precise ignition mechanism, the results presented in
this paper strongly suggest that explosions of sub-\mch\ WDs are a
viable scenario for fast-declining \sneia\ at the faint end of the
WLR, whose colours cannot be reproduced by \mch\ models.
In several upcoming companion papers we will present an in-depth study
of the sub-\mch\ model SCH2p0 compared to the low-luminosity
SN~1999by, and explore more generally the feasibility of such
sub-\mch\ models to reproduce the observed properties of more luminous
events, to address the question of multiple progenitor channels for
Type Ia supernovae.


\section*{Acknowledgements}

SB acknowledges useful discussions with Inma Dom\'inguez, Stuart Sim
and Marten van Kerkwijk. Part of this work was realized during a
one-month visit of SB to ESO as part of the ESO Scientific Visitor
Programme.  LD and SB acknowledge financial support from the Programme
National de Physique Stellaire (PNPS) of CNRS/INSU, France. DJH
acknowledges support from STScI theory grant HST-AR-12640.01, and NASA
theory grant NNX14AB41G. This work was granted access to the HPC
resources of CINES under the allocation c2014046608 made by GENCI
(Grand Equipement National de Calcul Intensif).  This work also
utilized computing resources of the mesocentre SIGAMM, hosted by the
Observatoire de la C\^ote d'Azur, Nice, France.



\bibliographystyle{mnras}
\bibliography{ms_wlr,atomic}



\appendix

\section{Model atoms}\label{sect:atom}

Table~\ref{tab_atom_big} gives the number of levels (both super-levels
and full levels; see \citealt{HM98} and \citealt{DH10} for details)
for the model atoms used in the radiative-transfer calculations
presented in this paper.

Oscillator strengths for CNO elements were originally taken from
\citet{NS83_LTDR, NS84_CNO_LTDR}. These authors also provide
transition probabilities to states in the ion continuum. The largest
source of oscillator data is from \citet{Kur09_ATD}\footnote{Data are
  available online at \url{http://kurucz.harvard.edu}}; its principal
advantage over many other sources (e.g., Opacity Project) is that LS
coupling is not assumed. More recently, non-LS oscillator strengths
have become available through the Iron Project \citep{HBE93_IP}, and
work done by the atomic-data group at Ohio State University
\citep{Nahar_OSU}. Other important sources of radiative data for Fe
include \citet{BB92_FeV, BB95_FeVI, BB95_FeIV}, \cite{Nahar95_FeII}.
Energy levels have generally been obtained from the National Institute
of Standards and Technology.  Collisional data is sparse, particularly
for states far from the ground state. The principal source for
collisional data among low lying states for a variety of species is
the tabulation by \citet{Men83_col}; other sources include
\citet{BBD85_col}, \citet{LDH85_CII_col}, \citet{LB94_N2},
\citet{SL74}, \citet{T97_SII_col,T97_SIII_col}, Zhang \& Pradhan
(\citeyear{ZP95_FeII_col,ZP95_FeIII_col,ZP97_FeIV_col}).
Photoionization data is taken from the Opacity Project
\citep{Sea87_OP} and the Iron Project \citep{HBE93_IP}. Unfortunately
Ni and Co photoionization data is generally unavailable, and we have
utilized crude approximations. Charge exchange cross-sections are from
the tabulation by \citet{KF96_chg}.

\begin{table}
\begin{center}
\caption[]{Summary of the model atoms used in our radiative-transfer
  calculations \citep[see also][]{DH10}.
  N$_{\rm f}$ refers to the number of full levels,
  N$_{\rm s}$ to the number of super levels, and N$_{\rm trans}$ to
  the corresponding number of bound-bound transitions. The last column
  refers to the upper level for each ion treated.
  Note that $n$w\,$^2$W refers to a
  state with principal quantum number $n$ (all $l$ states combined
  into a single state), and spin 2.
\label{tab_atom_big}}
\begin{tabular}{l@{\hspace{3mm}}r@{\hspace{3mm}}r@{\hspace{3mm}}r@{\hspace{3mm}}l}
\hline
 Species        &  N$_{\rm f}$  &  N$_{\rm s}$ & N$_{\rm trans}$ & Upper Level \\
\hline
     C{\,\sc i}      &  26  &   14 &    120 & 2s2p$^3$\,$^3$P\opar                  \\
     C{\,\sc ii}     &  26  &   14 &     98 & 2s$^2$4d\,$^2$D$_{5/2}$                \\
     C{\,\sc iii}    & 112  &   62 &    906 & 2s8f\,$^1$F\opar                       \\
     O{\,\sc i}      & 101  &   38 &    743 & 2s$^2$2p$^3$($^4$S\opar)6f\,$^3$F$_{2}$         \\
     O{\,\sc ii}     & 143  &   63 &   1868 & 2s$^2$2p$^2$($^3$P)5p\,$^2$P\oparsub{3/2}       \\
     O{\,\sc iii}    &  86  &   44 &    516 & 2s2p$^2$($^4$P)3p\,$^3$P\opar                   \\
     Ne{\,\sc i}     & 139  &   70 &   1587 & 2s$^2$2p$^5$($^2$P\oparsub{3/2})6d\,$^2$[5/2]\oparsub{3}  \\
     Ne{\,\sc ii}    &  91  &   22 &   1107 & 2s$^2$2p$^4$($^3$P)4d\,$^2$P$_{3/2}$       \\
     Ne{\,\sc iii}   &  56  &   24 &    240 & 2s$^2$2p$^3$($^3$S\opar)4p\,$^5$P  \\
     Na{\,\sc i}     &  71  &   22 &   1614 & 30w\,$^2$W                         \\
     Mg{\,\sc ii}    &  80  &   31 &   1993 & 30w\,$^2$W                         \\
     Mg{\,\sc iii}   &  99  &   31 &    775 & 2p$^5$7s\,$^1$P\opar                   \\
     Al{\,\sc ii}    &  44  &   26 &    171 & 3s5d\,$^1$D$_{2}$                    \\
     Al{\,\sc iii}   &  80  &   31 &   2011 & 30w\,$^2$W                         \\
     Si{\,\sc ii}    &  59  &   31 &    354 & 3s$^2$7g\,$^2$G$_{7/2}$             \\
     Si{\,\sc iii}   &  61  &   33 &    312 & 3s5g\,$^3$G$_{5}$                    \\
     Si{\,\sc iv}    &  48  &   37 &    405 & 10f\,$^2$F\opar                       \\
     S{\,\sc ii}     & 324  &   56 &   8464 & 3s3p$^3$($^5$S\opar)4p\,$^6$P             \\
     S{\,\sc iii}    &  98  &   48 &    840 & 3s3p$^2$($^2$D)3d$\,^3$P             \\
     S{\,\sc iv}     &  67  &   27 &    397 & 3s3p($^3$P\opar)4p\,$^2$D$_{5/2}$         \\
     Ar{\,\sc i}     & 110  &   56 &   1541 & 3s$^2$3p$^5$($^2$P\oparsub{3/2})7p\,$^2$[3/2]$_2$  \\
     Ar{\,\sc ii}    & 415  &  134 &  20197 & 3s$^2$3p$^4$($^3$P$_1$)7i\,$^2$[6]$_{11/2}$  \\
     Ar{\,\sc iii}   & 346  &   32 &   6901 & 3s$^2$3p$^3$($^2$D\opar)8s\,$^1$D\opar\            \\     
     Ca{\,\sc ii}    &  77  &   21 &   1736 & 3p$^6$30w\,$^2$W                     \\
     Ca{\,\sc iii}   &  40  &   16 &    108 & 3s$^2$3p$^5$5s\,$^1$P\opar                \\
     Ca{\,\sc iv}    &  69  &   18 &    335 & 3s3p$^5$($^3$P\opar)3d\,$^4$D\oparsub{1/2}        \\
     Sc{\,\sc ii}    &  85  &   38 &    979 & 3p$^6$3d4f$\,^1$P\oparsub{1}             \\
     Sc{\,\sc iii}   &  45  &   25 &    235 & 7h\,$^2$H\oparsub{11/2}                  \\
     Ti{\,\sc ii}    & 152  &   37 &   3134 & 3d$^2$($^3$F)5p\,$^4$D\oparsub{7/2}            \\
     Ti{\,\sc iii}   & 206  &   33 &   4735 & 3d6f\,$^3$H\oparsub{6}                   \\
     Cr{\,\sc ii}    & 196  &   28 &   4162 & 3d$^4$($^3$G)4p\,x$^4$G\oparsub{11/2}          \\
     Cr{\,\sc iii}   & 145  &   30 &   2359 & 3d$^3$($^2$D$_2$)4p\,$^3$D\oparsub{3}             \\
     Cr{\,\sc iv}    & 234  &   29 &   6354 & 3d$^2$($^3$P)5p\,$^4$P\oparsub{5/2}            \\
     Mn{\,\sc ii}    &  97  &   25 &    236 & 3d$^4$($^5$D)4s$^2$\,c$^5$D$_{4}$            \\
     Mn{\,\sc iii}   & 175  &   30 &   3173 & 3d$^4$($^3$G)4p\,y$^4$H\oparsub{13/2}         \\
     Fe{\,\sc i}     & 136  &   44 &   1900 & 3d$^6$($^5$D)4s4p\,x$^5$F\oparsub{3}           \\
     Fe{\,\sc ii}    & 827  &  275 &44\,831 & 3d$^5$($^6$S)4p$^2$($^3$P)\,$^4$P$_{1/2}$        \\
     Fe{\,\sc iii}   & 698  &   83 &36\,807 & 3d$^5$($^2$H)4d$^1$K$_{7}$               \\
     Fe{\,\sc iv}    &1000  &  100 &72\,223 & 3d$^4$($^3$G)4f\,$^4$P\oparsub{5/2}            \\
     Fe{\,\sc v}     & 191  &   47 &   3977 & 3d$^3$($^4$F)4d\,$^5$F$_{3}$              \\
     Fe{\,\sc vi}    & 433  &   44 &14\,103 & 3p$^5$($^2$P)3d$^4$($^1$S)\,$^2$P\oparsub{3/2}      \\
     Fe{\,\sc vii}   & 153  &   29 &   1753 & 3p$^5$($^2$P)3d$^3$($^2$D)\,$^1$P\oparsub{1} \\ 
     Co{\,\sc ii}    &2747  &  136 &593\,559& 3d$^7$($^2$D)6p\,$^3$P\oparsub{1}             \\
     Co{\,\sc iii}   &3917  &  123 &679\,412& 3d$^6$($^3$D)6d\,$^4$P$_{3/2}$            \\
     Co{\,\sc iv}    & 314  &   37 &   9062 & 3d$^5$($^2$P)4p\,$^3$P\oparsub{1}              \\
     Co{\,\sc v}     & 387  &   32 &13\,605 & 3d$^4$($^3$F)4d\,$^2$H$_{9/2}$            \\
     Co{\,\sc vi}    & 323  &   28 &   9608 & 3d$^3$($^2$D)4d$\,^1$S \\
     Co{\,\sc vii}   & 319  &   31 &   9096 & 3p$^5$($^2$P)4d($^3$F)\,$^2$D\oparsub{3/2} \\
     Ni{\,\sc ii}    &1000  &   59 &  51812 & 3d$^8$($^3$F)7f\,$^4$I\oparsub{9/2}            \\
     Ni{\,\sc iii}   &1000  &   42 &  66511 & 3d$^7$($^2$D)4d\,$^3$S$_{1}$             \\
     Ni{\,\sc iv}    & 254  &   28 &   6317 & 3D$^6$($^1$G$_1$)4p\,$^2$G\oparsub{7/2}           \\
     Ni{\,\sc v}     & 183  &   46 &   3065 & 3d$^5$($^2$D3)4p\,$^3$F\oparsub{3}            \\
     Ni{\,\sc vi}    & 314  &   37 &   9569 & 3d$^4$($^5$D)4d\,$^4$F$_{9/2}$            \\
     Ni{\,\sc vii}   & 308  &   37 &   9225 & 3d$^3$($^2$D)4d\,$^3$P$_{2}$ \\
\hline
     {\bf Total}     & {\bf 18\,710} & {\bf 2534} & {\bf 1\,717\,141} & \\
\hline
\end{tabular}
\end{center}
\end{table}

\section{Contribution of individual ions to the total optical
  flux}\label{sect:ladder} 

\begin{figure*}
\centering
\includegraphics{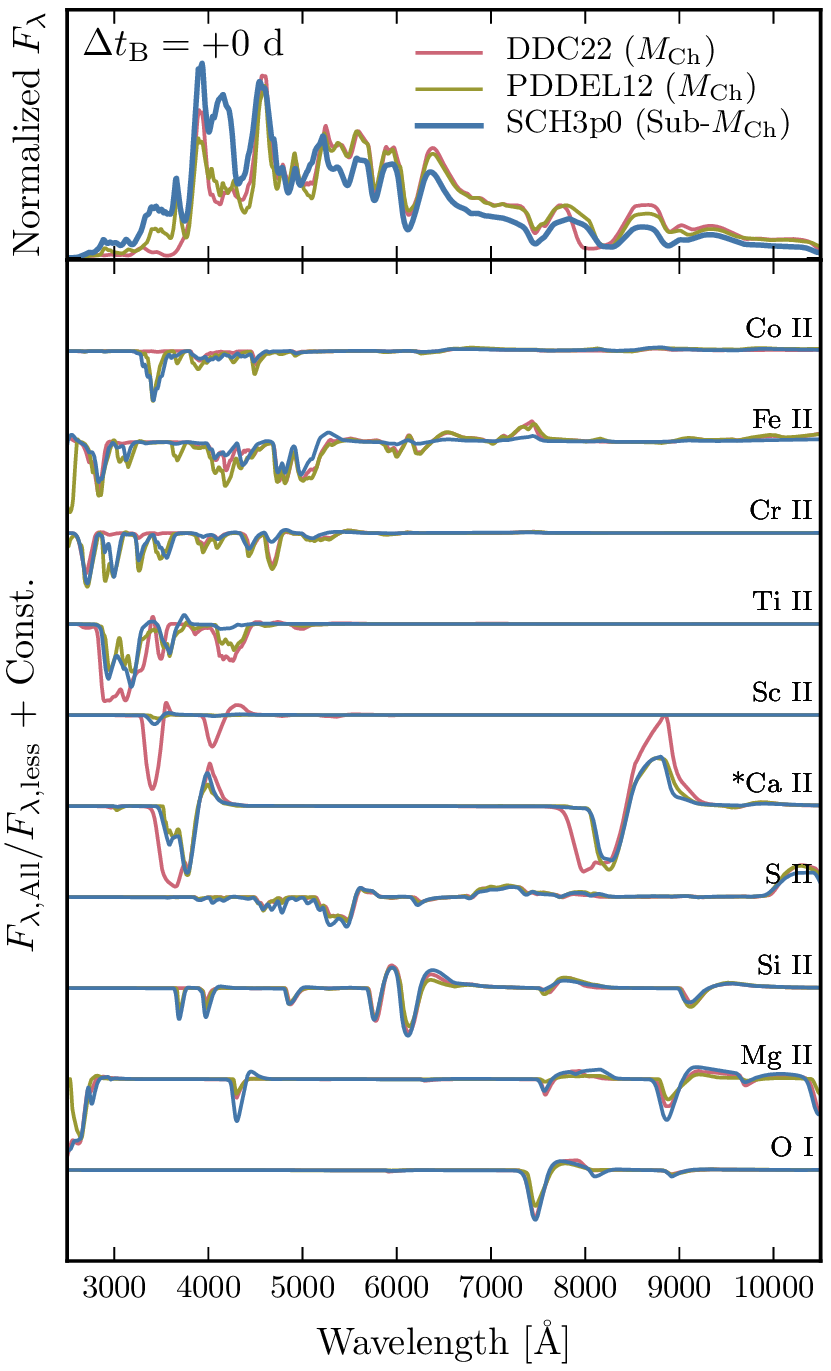}\hspace{.5cm}
\includegraphics{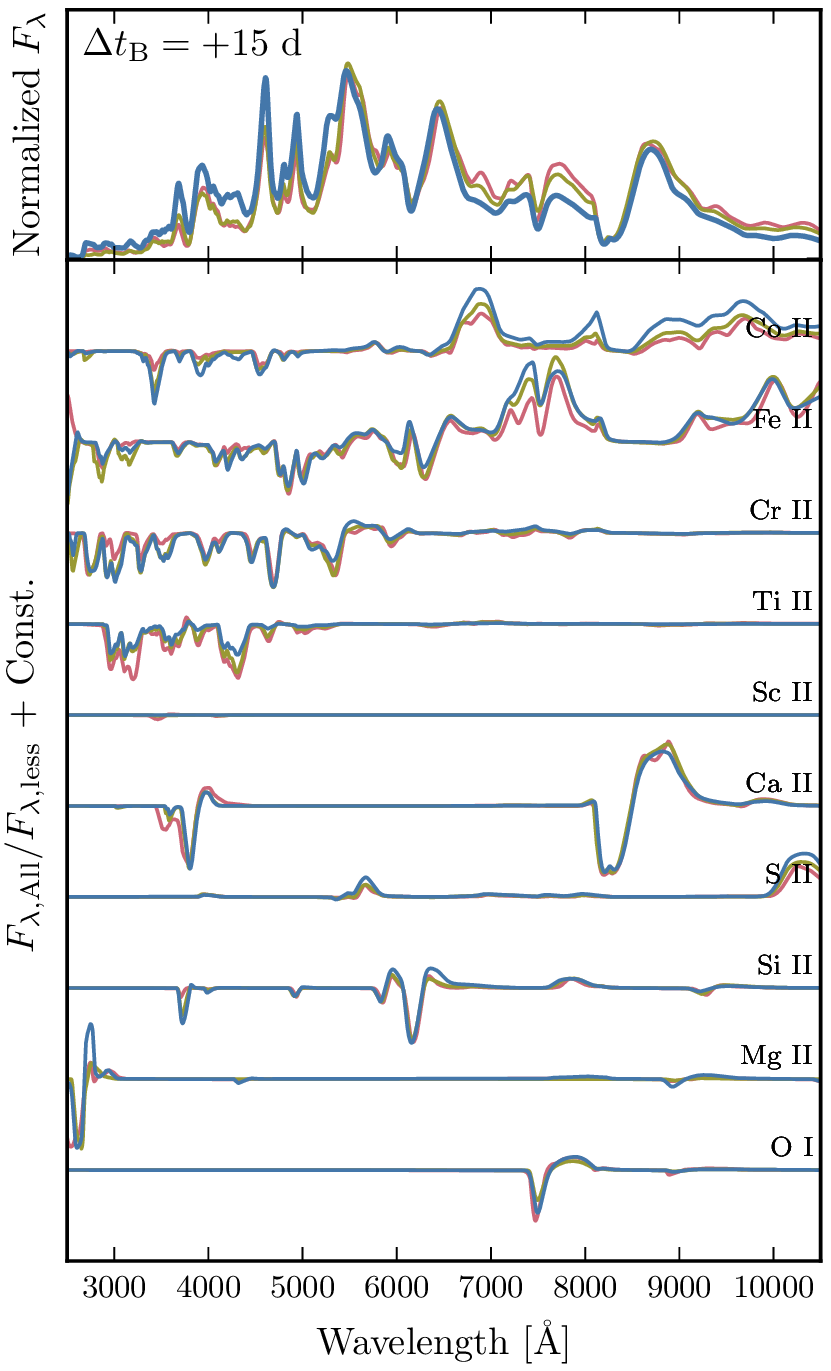}
\caption{\label{fig:ladderp0p15}
Contribution of individual ions (bottom panels) to the full optical
synthetic spectra of the standard \mch\ delayed-detonation model DDC22
(red line), the pulsational \mch\ delayed-detonation model PDDEL12
(yellow line), and the sub-\mch\ model SCH3p0 ($M_{\rm
  tot}=0.95\,\msun$; blue line), all of which have a similar
\nifs\ mass of $0.23\pm0.02$\,\msun, at $B$-band maximum light (left)
and 15 days later (right). The spectra have been normalized to their
mean flux in the range 3000--10000\,\AA.  Only ions that impact the
flux at the $>10$ per cent level at either phase are shown.  The
Ca\two\ ion spectra marked with a ``*'' have been scaled down for
clarity.
}
\end{figure*}

Figure~\ref{fig:ladderp0p15} reveals the contribution of individual
ions (bottom panels) to the full optical synthetic spectra of the
standard \mch\ delayed-detonation model DDC22, the pulsational
\mch\ delayed-detonation model PDDEL12, and the sub-\mch\ model SCH3p0
($M_{\rm tot}=0.95\,\msun$), all of which have a similar \nifs\ mass
of $0.23\pm0.02$\,\msun, at $B$-band maximum light and 15 days
later. Only ions that impact the flux at the $>10$
per cent level at either phase are shown.

\bsp	
\label{lastpage}
\end{document}